\documentclass{article}
\usepackage{spconf,amsmath,graphicx}

\usepackage{amsmath,amssymb,amsthm,amsfonts,latexsym,bbm,xspace,graphicx,float,mathtools,paralist,
verbatim, xcolor} 
\usepackage{url}
\PassOptionsToPackage{hyphens}{url}\usepackage{hyperref}

\newcommand{\probe}{\alpha}
\newcommand{\response}{\beta}
\newcommand{\utility}{u}

\newcommand{\reals}{\mathbb{R}}
\newcommand{\dataset}{\mathcal{D}}
\newcommand{\statenoisecov}{Q}
\newcommand{\obsnoisecov}{R}
\renewcommand{\time}{k}
\newcommand{\horizon}{K}
\newcommand{\sigdim}{m}

\newcommand{\argmin}[1]{\operatorname{argmin}_{#1}}
\newcommand{\argmax}[1]{\operatorname{argmax}_{#1}}

\newtheorem{definition}{Definition}
\newtheorem{theorem}{Theorem}

\title{How can a Cognitive Radar Mask its Cognition?}
\name{Kunal Pattanayak \sthanks{V. Krishnamurthy and K. Pattanayak are with the School	of Electrical and Computer Engineering, Cornell University, Ithaca,	NY, 14853 USA. e-mail: vikramk@cornell.edu, kp487@cornell.edu.
		$^\dag$ C. Berry is with Lockheed Martin Advanced Technology Laboratories, Cherry Hill, NJ, 08002 USA. e-mail: christopher.m.berry@lmco.com. This research was supported in part by a research contract from  Lockheed Martin and  the Army Research Office grant W911NF-21-1-0093.}, Vikram Krishnamurthy$^\ast$ and Christopher Berry$^\dagger$}
\address{$^\ast$ Electrical and Computer Engineering, Cornell University, USA\\
$^\dag$ Lockheed Martin Advanced Technology Laboratories, USA. }

\begin{document}
%
\maketitle
\begin{abstract}
We study how a cognitive radar can mask (hide) its  cognitive ability from an adversarial jamming device.  Specifically, if the radar optimally adapts its waveform based on adversarial target maneuvers (probes), how should the radar choose its waveform parameters (response) so that its utility function cannot be recovered by the adversary. This paper abstracts the radar's cognition masking problem in terms of the spectra (eigenvalues) of the state and observation noise covariance matrices, and embeds the algebraic Riccati equation into an economics-based utility maximization setup. Given an observed sequence of radar responses, the adversary tests for utility maximization behavior of the radar and estimates its utility function that rationalizes the radar's responses. In turn, the radar deliberately chooses sub-optimal responses so that its utility function almost fails the utility maximization test, and hence, its cognitive ability is masked from the adversary. We illustrate the performance of our cognition masking scheme via simple numerical examples. Our approach in this paper is based on revealed preference theory in microeconomics for identifying rationality.

\end{abstract}
\begin{keywords}
Cognitive Radar, Revealed Preference, Adversarial Inverse Reinforcement Learning, Electronic Counter Countermeasures, Kalman Filter
\end{keywords}
\section{Introduction}

In abstract terms, a cognitive radar is a utility maximizer - it adapts its waveform, scheduling and beam by optimizing  utility functions.
Consider the scenario where an adversarial target  probes a cognitive radar (to possibly degrade the radar's performance) and analyzes the radar's responses to estimate the radar's  utility function. {\em How can  the radar covertly mask its utility function by deliberately choosing responses that confuse the adversary?} In this paper, we propose a revealed preference-based approach to mask the radar's cognition  with the working assumption that the cognitive radar satisfies economics-based rationality.
\begin{figure}[h]
    \centering
    \includegraphics[width=0.85\columnwidth]{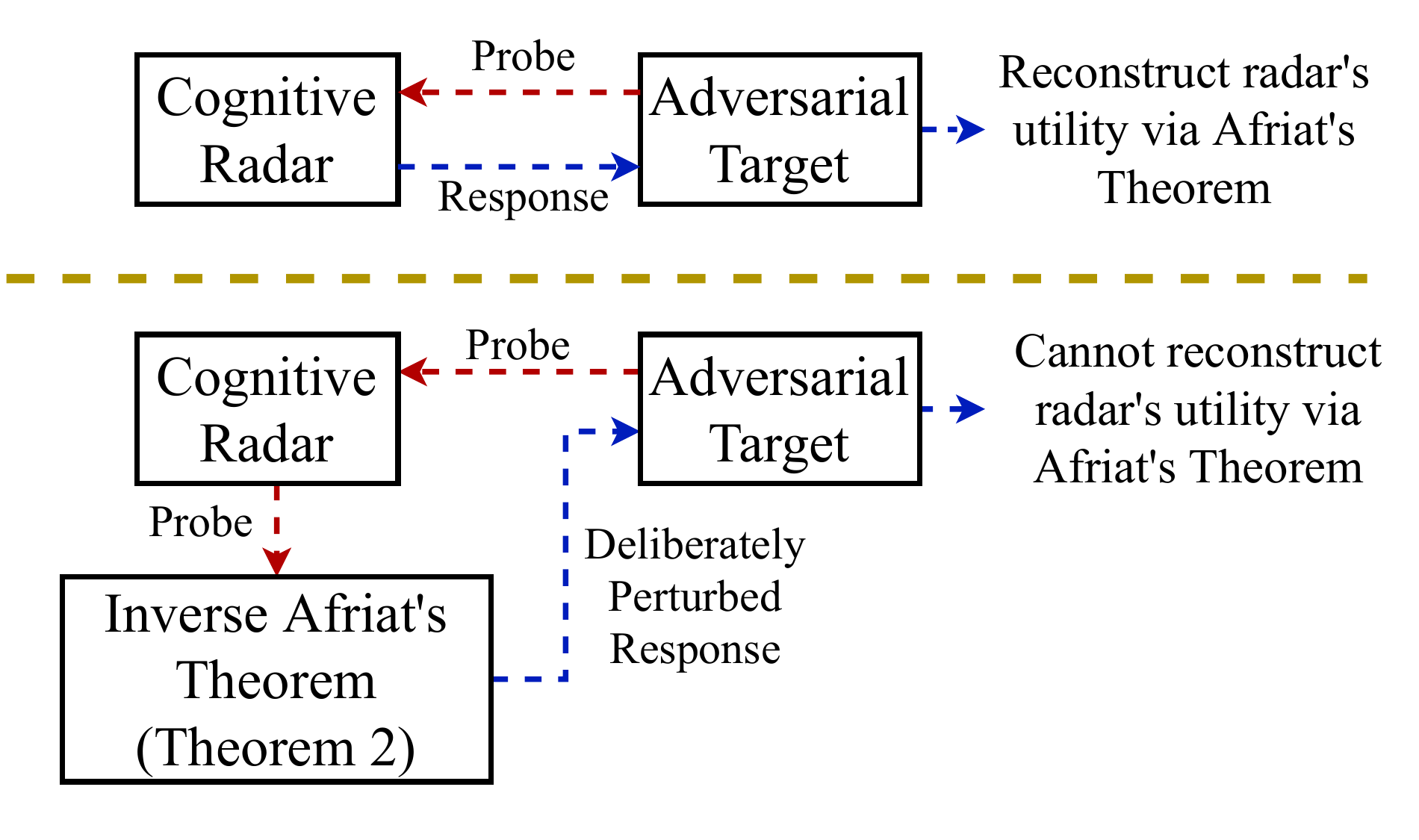}
    \caption{Schematic of the Masking Cognition problem. The adversarial target sends a sequence of probe signals to the radar and records its responses. If the cognitive radar responds naively to the adversarial target's probes, its utility function can be recovered via Afriat's theorem (Top). If the radar deliberately perturbs its response using the {\em inverse} Afriat's theorem, the adversary fails to reconstruct its utility (Bottom).}
    \label{fig:schematic}
\end{figure}

Before going into the details, we emphasize that the problem formulation and algorithms  developed here also apply to {\em adversarial inverse reinforcement learning}. In inverse reinforcement learning \cite{NR00,ABB04,ZB08,KY20}, an inverse learner seeks to estimate the utility function of a decision maker by observing its decisions. A natural extension is: How can the decision maker hide its utility function by slightly perturbing the actions it takes in the presence of an adversary? 

\noindent {\bf Related Work.}
This paper builds on our previous work \cite{KAEM20} where  the cognitive radar is not aware that it is being probed by the adversarial target. 
If the radar is aware of the adversary's motives, how to deliberately respond with sub-optimal responses to confuse the adversarial target (see Fig.\,\ref{fig:schematic})?
In a companion paper \cite{GK21} we formulate the electronic counter countermeasure problem as a principal agent problem where the radar and adversary establish an information asymmetric contract. In comparison, the formulation in the current paper is adversarial where the radar seeks to confuse the adversary.

This paper can be viewed in the context of low-probability of intercept (LPI) radar as a countermeasure to electronic intelligence (ELINT) gathering targets~\cite{W06}. Masking cognition in the face of an adversarial target is closely related to the areas of electronic counter countermeasures (ECCM) and RF stealth~\cite{stealthmain} in electronic warfare. \cite{ECCMsurvey} provides a comprehensive list of ECCM techniques. \cite{ECCM1,ECCM2} propose waveform adaptation schemes to counter barrage jamming. \cite{ECCM3} proposes time-frequency based ECCM solutions for deceptive
jamming. \cite{stealth1,stealth2,stealth3} exploit frequency diversity for radio stealth in multi-target and moving target tracking. However, cognition masking strategies with minimal performance loss have not been explored previously.

\subsubsection*{Background. Revealed Preference and Afriat's Theorem}
Our approach to masking cognition in radars is based on revealed preference in micro-economics. The area of revealed preference~\cite{Sam38,Afr67,Die73,Var82} focuses on nonparametric detection of utility maximization behavior given a finite dataset of probe and response signals.

\begin{definition}\label{def:UM}
A system is a utility maximizer if for a probe signal $\probe\in\reals_+^m$, the response signal $\response\in\reals_+^m$ satisfies:
\begin{equation} \label{eq:utilitymaximization}
    \response = \argmax{\bar{\response}\in\reals_+^m} u(\bar \response),~\probe'\bar \response\leq 1,
\end{equation}
where $u$ is a monotone utility function.
\end{definition}
In micro-economics, probe $\probe$ is the vector of prices of a set of goods, and response $\response$ is the consumption vector. Hence, the constraint $\probe'\bar{\response}\leq 1$ is a budget constraint with total budget $\$1$. Indeed, this constraint can be replaced WLOG by $\probe'\bar{\response}\leq c$, where $c>0$ is the actual budget. Given a finite time series of probes and responses from a system, how to test is the system is a utility maximizer \eqref{eq:utilitymaximization}? 

The key result in revealed preference is Afriat's theorem \cite{Afr67,Die73}. A remarkable property of Afriat's theorem is that it gives testable conditions that are both necessary and sufficient conditions for a time series of probes and responses to be consistent with utility maximization behavior~\eqref{eq:utilitymaximization}. 
\begin{theorem}[Afriat's Theorem~\cite{Afr67}] Given a sequence of probes and responses $\dataset=\{(\probe_\time,\response_\time), \time\in \{1,2,\dots,\horizon\}\}$,
 the following statements are equivalent:
	\begin{compactenum}
	\item There exists a monotone, continuous and concave utility function that satisfies \eqref{eq:utilitymaximization}. 
		\item \underline{\em Afriat's Test:} There exist reals $u_t,\lambda_t>0,~t=1,2,\ldots,\horizon,$ such that the following inequalities are feasible.
			\begin{equation}
				u_s-u_t-\lambda_t \probe_t' (\response_s-\response_t) \leq 0 \; \forall t,s\in\{1,\dots,\horizon\}.\
				\label{eqn:AfriatFeasibilityTest}
			\end{equation}
			The monotone, concave utility function
			given by 
			\begin{equation}
				\utility(\response) = \underset{t\in \{1,2,\dots,\horizon\}}{\operatorname{min}}\{u_t+\lambda_t \probe_t'(\response-\response_t)\}
				\label{eqn:estutility}
            \end{equation}
            constructed using $u_t$ and $\lambda_t$ \eqref{eqn:AfriatFeasibilityTest} rationalizes $\mathcal{D}$ \eqref{eq:utilitymaximization}.
          \item The data set $\mathcal{D}$ satisfies the Generalized Axiom of Revealed Preference (GARP), 
          namely for any $t \leq \horizon$,
           $\probe_t' \response_t \geq \probe_t' \response_{t+1} \quad \forall t\leq k-1 \implies \probe_k  \response_k \leq \probe_k'  \response_{1}$.
	\end{compactenum}
\label{thrm:rp}
\end{theorem}
Afriat's theorem tests for economics-based rationality. In the radar context, the adversarial target uses Afriat's theorem to test for the radar's cognition. If Afriat's inequalities \eqref{eqn:AfriatFeasibilityTest} have a feasible solution, then the adversary constructs a set of feasible utility functions \eqref{eqn:estutility} that rationalize the radar's responses. The estimated utility is set-valued since the reconstructed utility is ordinal - any positive monotone transformation of a feasible utility function rationalizes the radar's responses. 

{\em Outline:} Sec.\,\ref{sec:background} below reconciles the abstract utility maximization setup of Definition~\ref{def:UM} with the radar's cognitive behavior, specifically, waveform adaptation during target tracking. Sec.\,\ref{sec:iirl} proposes a cognition masking strategy for the radar when the radar knows an adversarial target is reconstructing its utility function. Finally, Sec.\,\ref{sec:numerical} illustrates the performance of the cognition masking scheme via  two numerical examples.

\section{Optimal Waveform Adaptation as Utility Maximization} \label{sec:background}
Waveform adaptation is a crucial functionality of a cognitive radar. In this section, we abstract optimal waveform adaptation of a cognitive radar using a Kalman filter for target tracking into the utility maximization setup of Definition~\ref{def:UM}. The key idea is to express the linear budget constraint of Definition~\ref{def:UM} in terms of the eigenvalues (spectra) of the state and observation noise covariances of the radar's state space model.

Linear Gaussian dynamics for a target’s kinematics~\cite{LJ03} and linear Gaussian measurements at the radar are widely assumed as a useful approximation~\cite{BLK08}. Accordingly, consider the following state space model for the radar:
\begin{equation}\begin{split}
x_{n+1} &= A x_n + w_n(\probe_\time), \quad x_0 \sim \pi_0
\label{eq:kalman-sys}\\
y_n &= C x_n + v_n(\response_\time) ,
\end{split}\end{equation}
where $x_n \in \mathcal{X} = \mathbb{R}^X$ is the target state with initial density  $\pi_0 \sim \mathcal{N}(\hat{x}_0, \Sigma_0)$,
$y_n \in \mathcal{Y} = \mathbb{R}^Y$ is the radar's observation,
$w_n \sim \mathcal{N}(0, Q(\probe_\time))$ and $v_n \sim \mathcal{N}(0, R(\response_\time))$ are mutually independent, Gaussian noise processes. 

The state noise covariance $Q$ is parameterized by the adversarial target's probe $\probe_\time$ and the observation noise covariance $R$ is parameterized by the radar's response $\response_\time$ (see \cite[Sec.\,III-B]{KAEM20} for a detailed discussion on the relation between radar's waveform and observation noise covariance $R$). 
It is important to distinguish between the subscripts $n,k$ in \eqref{eq:kalman-sys}. The subscript $n$ indicates system updates at the tracker level (faster timescale), and the subscript $\time$ indicates the epoch (slower timescale) for the probe and response. When state $x_n$ represents the position and velocity in Euclidean space, $A$ is a block diagonal constant velocity matrix \cite{BP99}.
The state noise covariance $Q(\probe)$ in \eqref{eq:kalman-sys} models acceleration maneuvers of the target parameterized by the probes $\probe$.

The radar estimates the target state $\hat{x}_n$ with covariance $\Sigma_n$ from observations $y_{1:n}$.
The posterior $\pi_n$ is propagated recursively in time via the classical Kalman filter equations:
\begin{align*}
\Sigma_{n+1|n} &= A \Sigma_n A' + Q(\probe_\time),~K_{n+1} = C \Sigma_{n+1|n} C' + R(\response_\time)\\
\psi_{n+1} &= \Sigma_{n+1|n} C' K^{-1}_{n+1}, ~\hat{x}_{n+1} = A \hat{x}_n + \psi_{n+1} (y_{n+1} - C A \hat{x}_n)\\
\Sigma_{n+1} &= (I - \psi_{n+1} C)\Sigma_{n+1|n}.
\end{align*}
Assuming the model parameters \eqref{eq:kalman-sys} satisfy the conditions that $[A, C]$ is detectable and $[A, \sqrt{Q}]$ is stabilizable,
the steady-state predicted covariance $\Sigma_\infty$ is the unique positive semi-definite solution of the \textit{algebraic Riccati equation} (ARE):
\begin{align}
\mathcal{A}(\probe_\time, \response_\time, \Sigma) = & 
-\Sigma + A(\Sigma -
 \Sigma C' [C \Sigma C' + R(\response)]^{-1} C \Sigma) A' \nonumber\\
 & \quad \quad + Q(\probe) = 0.\label{eqn:ARE}
\end{align}
Denote $\Sigma^*(\probe_\time, \response_\time)$ as the solution of the ARE given probe $\probe_\time$ and response $\response_\time$ at time $k$. 

Our working assumption is that the radar maximizes a utility function $u$ to choose its optimal waveform at the start of every epoch $\time$. Hence, it only remains to justify the linear budget $\probe_\time'\response_\time\leq 1$ in Definition~\ref{def:UM} to embed waveform optimization into the utility maximization setup. We suppose:
\begin{compactitem}
    \item the target probe $\probe$ is the vector of eigenvalues of the positive definite matrix $Q$
    \item the radar response $\response$ is the vector of eigenvalues of the positive definite matrix $R^{-1}$.
\end{compactitem}

The $i^{\text{th}}$ component of $\response_\time$ is the measurement precision (amount of energy) of the radar in the $i^{\text{th}}$ mode. Similarly, the $i^{\text{th}}$ component of $\probe_\time$ is the incentive for considering the $i^{\text{th}}$ mode of the target. Put together, $\probe_\time'\response_\time$ measures the signal-to-noise ratio (SNR) of the radar. Thus, $\probe_\time'\response_\time\leq 1$ is effectively a bound on the radar's SNR. Hence, in the utility maximization context, the radar chooses the most precise observation noise covariance $R(\response_\time)$ such that its SNR lies below a particular threshold \footnote{see \cite{KAEM20} for a more detailed discussion on the linear budget in terms of the solution to the ARE \eqref{eqn:ARE}.}.

To summarize, we have justified how the cognitive radar's waveform adaptation can be cast as the constrained utility maximization problem of Definition~\ref{thrm:rp}. Hence, the adversarial target can now use Afriat's Theorem~\ref{thrm:rp} to reconstruct the radar's utility. How should the radar react so that its utility function is be recovered accurately? The rest of the paper focuses on a cognition masking strategy for the radar. The key idea is for the radar to deliberately choose sub-optimal waveforms so that the radar's utility $u$ satisfies the Afriat inequalities \eqref{eqn:AfriatFeasibilityTest} by a small margin, thus confusing the adversarial target at the cost of performance degradation. 




\section{Inverse Revealed Preference for Masking Utility Function } \label{sec:iirl}
\begin{figure}
    \centering
    \includegraphics[width=0.65\columnwidth]{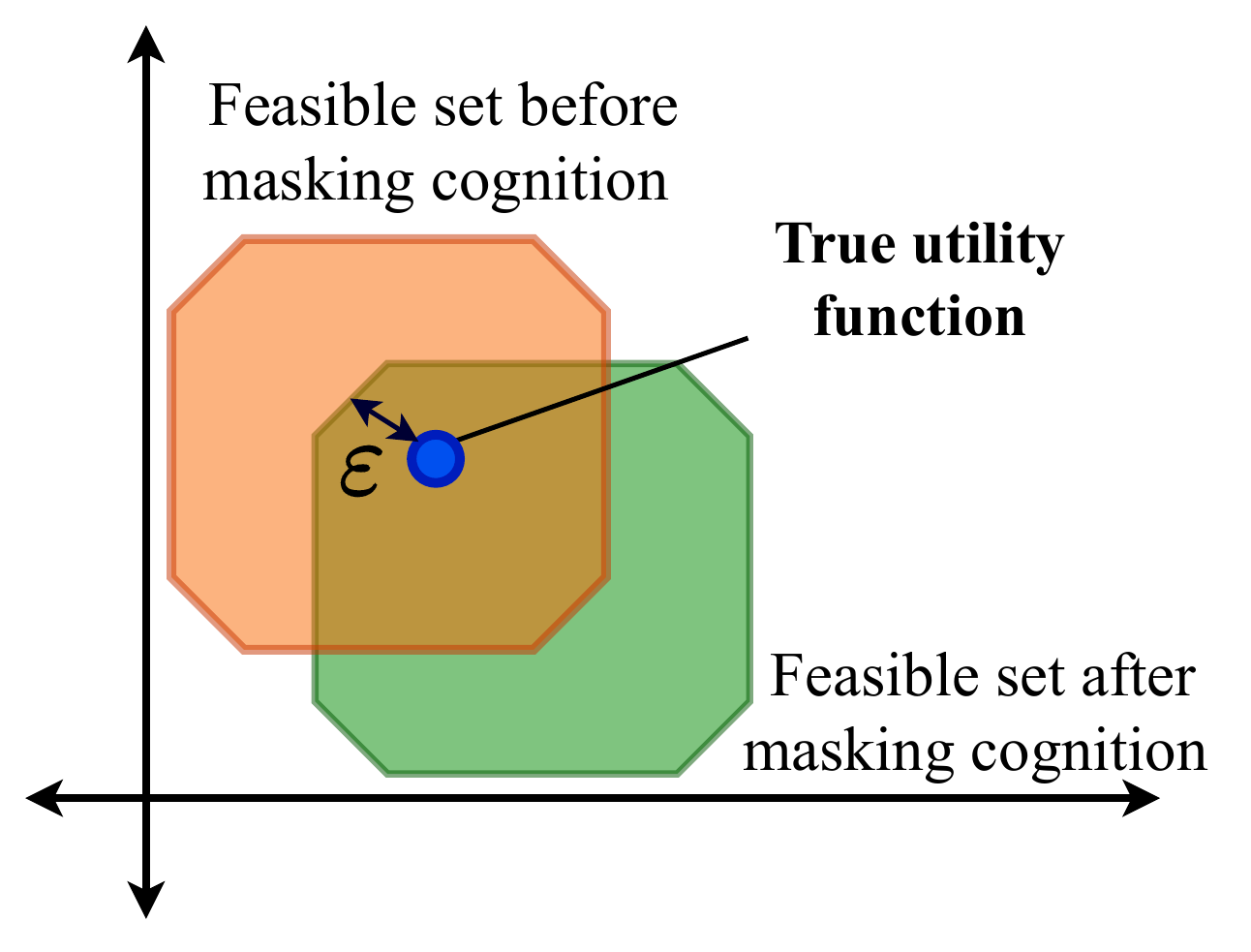}
    \caption{Masking Cognition by Performance Degradation.
    If the radar responds naively to the adversary target's probes, its utility passes the utility maximization test by a large margin and hence, is close to the center of the feasible set (orange region) computed by the adversary.
    By deliberately perturbing its response and degrading its performance, the radar shifts the feasible set (green region) so that the true utility is within $\epsilon$ distance from the edge of the set.}
    \label{fig:illth2}
\end{figure}
We now present the main result of this paper, namely, inverse Afriat's Theorem. If the adversary uses Afriat's theorem to reconstruct the radar's utility function, the radar uses the inverse Afriat Theorem below to deliberately perturb its responses and mask its utility function. Put differently, the radar deliberately compromises on its performance to prevent the reconstruction of its utility function.

In terms of the radar's choice of waveform parameters, the radar 
chooses sub-optimal sensing modes (observation noise covariance) given the adversarial target's maneuvers (state noise covariance) so that the radar's utility function is masked from the adversary.
\begin{theorem}[Inverse Afriat's Theorem to Mask Cognition]
Suppose the radar optimizes a monotone, continuous utility function $u$, and the adversary uses Theorem~\ref{thrm:rp} to estimate the radar's utility function. Given the adversary's probe sequence $\{\probe_\time\}_{\time=1}^{\horizon}$, the radar's response sequence $\{\response_\time\}_{\time=1}^{\horizon}$ that masks its utility function is given by: 
\begin{equation}
    \response_\time = \response_\time^\ast + \eta_\time^\ast. \label{eqn:perturbedresponse}
\end{equation}
In \eqref{eqn:perturbedresponse}, $\response_\time^\ast$ is the optimal response to the probe signal $\probe_\time$:
\begin{equation}\label{eqn:naiveresponse}
    \response_\time^\ast = \argmax{\response\in\mathbb{R}^m_+} u(\response),~\text{s.t. }\probe_\time'\response\leq 1.
\end{equation}
The sequence $\{\eta_\time^\ast\}_{\time=1}^\horizon$ is the minimum perturbation that ensures the radar's utility function passes the adversary's test for utility maximization \eqref{eqn:AfriatFeasibilityTest} by a margin less than $\epsilon\in\reals_+$: \vspace{-0.6cm}
\begin{align}
 \eta_{1:\horizon}^\ast &= \argmin{\eta_{1:\horizon}} \sum_{\time=1}^\horizon \|\eta_\time\|_2^2, \label{eqn:lowmarginmask}\\     u(\response_s) & \geq u(\response_t)  
 - \nabla u(\response_t)'(\response_s -\response_t) + \epsilon,~\forall~s,t\label{eqn:constraint_lowmargin}\\
\response_t^\ast + \eta_t& \geq 0,~\forall~t=1,2,\ldots,\horizon \label{eqn:constaint_nonnegative}
\end{align}
The variable $\epsilon>0$ is a user-defined parameter that determines the extent of cognition masking. 
\label{thrm:irp}
\end{theorem}

\begin{figure*}
    \centering
  \includegraphics[width=0.8\linewidth]{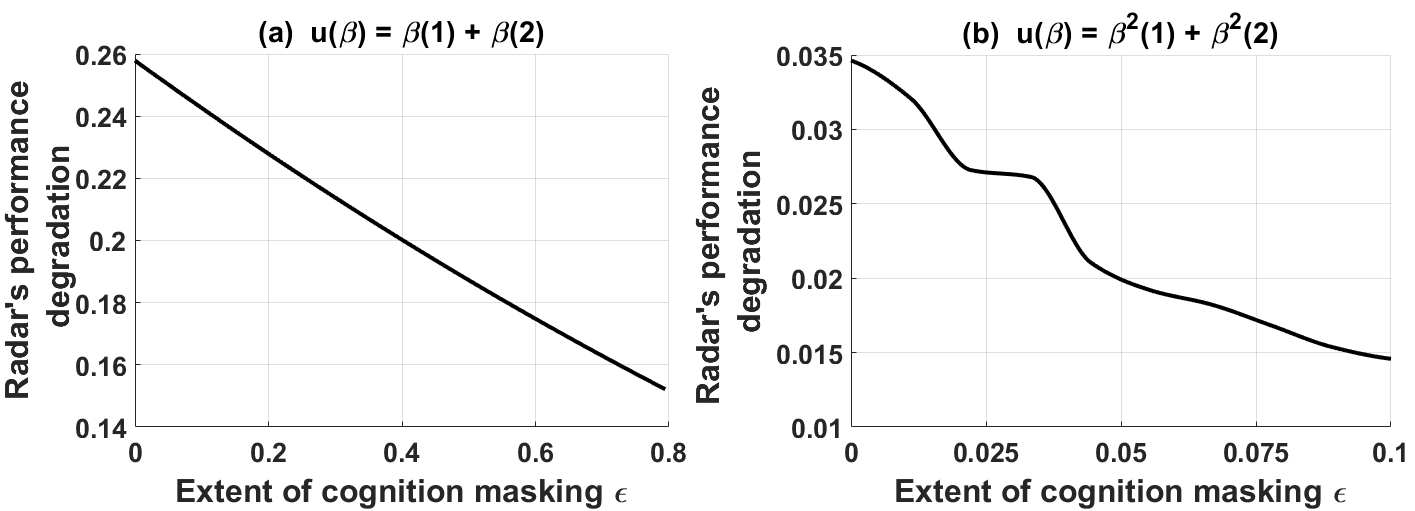}
    \caption{Deliberate performance loss (vertical-axis) of the cognitive radar \eqref{eqn:lowmarginmask} as a function of $\epsilon$ (horizontal-axis) which  measures the extent of cognition masking on the adversary's side \eqref{eqn:lowmarginmask}. (i) $\epsilon=0$ corresponds to maximum cognition masking and hence results in maximum performance loss. (ii) Due to larger local variation, for a fixed value of $\epsilon$, the quadratic utility (sub-figure (b)) requires smaller perturbation ($\approx$ 10 times) from the optimal response 
    compared to the linear utility (sub-figure (a)). }
    \label{fig:detconfplot}
\end{figure*}

Theorem~\ref{thrm:irp} masks the cognitive radar's utility function by deliberately perturbing its responses so that the responses almost fail the adversary's test for utility maximization (Theorem~\ref{thrm:rp}). If the radar naively responds to the adversary's probes ($\eta_\time^\ast=0,~\forall\time$), the radar's utility function passes the Afriat's test \eqref{eqn:AfriatFeasibilityTest} by a large margin and is thus a high-confidence utility estimate for the adversary. Since the radar's utility passes the Afriat's test by a very small margin due to the masking scheme of Theorem~\ref{thrm:irp}, it now lies very close to the edge of the feasible set of utilities\footnote{It follows from simple observation that utilities that pass Afriat's test with zero margin form the edge of the set of utilities for which Afriat's inequalities are feasible. Hence, the margin by which a utility function passes Afriat's test is proportional to its distance from the edge of the feasible set.}, and is no more a high-confidence utility estimate for the adversary. \vspace{0.15cm}

\noindent {\em Extent of cognition masking $\epsilon$.}
A smaller value of $\epsilon$ implies better cognition masking and higher performance degradation of the radar.
Setting $\epsilon$ to $0$ in \eqref{eqn:constraint_lowmargin} completely masks the radar's utility function ($u$ lies on the edge of the feasible set), but requires maximum degradation of radar performance (large perturbation \eqref{eqn:lowmarginmask} from the optimal response \eqref{eqn:naiveresponse}). On the other extreme, a large value of $\epsilon$ results in zero performance loss of the radar, but exposes the radar's utility function to the adversary since it lies very close to the center of the feasible set. 

\section{Numerical examples}\label{sec:numerical}
 Theorem~\ref{thrm:irp} specified the procedure for a cognitive radar to effectively mask its cognition from an adversarial target. Below, we illustrate via simple numerical examples the masking performance of Theorem~\ref{thrm:irp} for two different utility functions.

We chose $\horizon=50$ and  $\sigdim=2$, the dimension of adversarial target's probe and radar's response. The elements of the adversarial target's probe signals are generated randomly and independently over time as $\probe_k(i)\sim$ Unif$(0.2,2.5)$ for all $i=1,2$ and time $\time=1,2,\ldots,\horizon$,where Unif($a,b$) denotes uniform pdf with support $(a, b)$. Recall  that the probe signal $\probe_\time$ is the diagonal of the state noise covariance matrix: $\statenoisecov_\time=\operatorname{diag}[\probe_\time(1),\probe_\time(2)]$.

Given the probe sequence $\{\probe_\time,\time=1,2,\ldots,\horizon\}$, the cognitive radar chooses its response sequence $\{\response_\time,\time=1,2,\ldots,\horizon\}$ via \eqref{eqn:perturbedresponse} in Theorem~\ref{thrm:irp}. Recall from Sec.\,\ref{sec:background} that response $\response_\time$ is the diagonal of the inverse of radar's observation noise covariance matrix: $\obsnoisecov_\time^{-1} = \operatorname{diag}[\response_\time(1),\response_\time(2)]$ .
We generate two separate sequences of responses for the same probe sequence, but for two different utility functions \eqref{eqn:naiveresponse}:
\begin{align*}
    (a)&~u(\response) = \response(1)  + \response(2),\qquad (b)~u(\response) = \response^2(1) + \response^2(2)
\end{align*}
Figure~\ref{fig:detconfplot} shows the loss in performance (minimum perturbation from optimal response~\eqref{eqn:lowmarginmask}) of the cognitive radar as a function of $\epsilon$ (extent of cognition masking), for both choices of utility functions.
%
From Fig.\,\ref{fig:detconfplot}, we see that for both utility functions, the radar's performance decreases with increasing $\epsilon$ (larger extent of utility masking). This is expected since larger $\epsilon$ implies larger shift of the feasible set of utilities constructed by the adversarial target.



\section{Conclusion and Extensions}
This paper focuses on masking a radar's cognition when probed by an adversarial target. 
Our main result is Theorem~\ref{thrm:irp} that describes the radar's cognition masking strategy. The radar deliberately chooses sub-optimal responses at the cost of its performance, but prevents its utility function from being recovered by the adversary. 

Finally,
a useful  extension of this paper would be to study more general game-theoretic settings where even the adversary knows the radar is trying to mask its cognition. {\em How to detect play from the Nash equilibrium of a game between the radar and adversary?}

\bibliographystyle{unsrt_abbrv_custom}
\newpage \newpage \bibliography{refs}

\end{document}